\documentclass[aps,twocolumn,showpacs,preprintnumbers,amsmath,amssymb,superscriptaddress,prl,floatfix,10pt]{revtex4-2}

\usepackage{graphicx}
\usepackage{dcolumn}
\usepackage{bm}
\usepackage{graphics}
\usepackage{subfigure}
\usepackage{epsfig}
\usepackage{epstopdf}
\usepackage{xcolor}
\usepackage{multirow}
\usepackage{mathtools}
\usepackage{comment}
\usepackage{placeins}
\usepackage{orcidlink}
\usepackage{hyperref}
\usepackage[normalem]{ulem}
\usepackage{cancel}

\newcommand\bluesout{\bgroup\markoverwith{\textcolor{blue}{\rule[0.5ex]{2pt}{0.4pt}}}\ULon}
\hypersetup{colorlinks=true,citecolor=blue,linkcolor=blue,urlcolor=blue}

\allowdisplaybreaks

\begin{document}
\title{Magnetotransport of tomographic electrons in a Corbino disk}

\author{Nitay Ben-Shachar\,\orcidlink{0009-0002-2989-2295}}
\email{nshachar@caltech.edu}
\affiliation{Department of Physics, California Institute of Technology, Pasadena CA, 91125, USA}

\author{Johannes Hofmann\,\orcidlink{0000-0002-0667-2452}}
\email{johannes.hofmann@physics.gu.se}
\affiliation{Department of Physics, Gothenburg University, 41296 Gothenburg, Sweden}
\affiliation{Nordita, Stockholm University and KTH Royal Institute of Technology, 10691 Stockholm, Sweden}

\date{\today}

\begin{abstract}
    In clean electron gases at low-to-moderate temperatures, odd-parity modes of the Fermi surface are anomalously long-lived due to Pauli blocking, giving rise to ``tomographic transport'' that is not captured by a hydrodynamic model. Here we show that tomographic flow in a Corbino disk induces an extended boundary layer near electrodes with superballistic transport and enhanced slip velocity, which leads to a parametric enhancement of the quadratic  magnetoresistance coefficient. The enhancement depends explicitly on the electrode curvature, allowing its strength to be controlled by the device geometry. The magnetoresistance coefficient reveals three distinct  regimes as a function of magnetic field: a tomographic  regime at weak fields; a hydrodynamic regime at intermediate fields, reached when the cyclotron radius becomes comparable to a large odd-mode mean free path; and a conventional Ohmic regime at large fields, reached when the cyclotron radius becomes comparable to the short even-mode mean free path. The tomographic regime is characterized by an anomalous dependence of the magnetoresistance on temperature and density, which may account for recent experimentally observed anomalous scaling of the electron viscosity. 
\end{abstract}

\maketitle

\begin{figure}
    \centering
    \includegraphics[width=\linewidth]{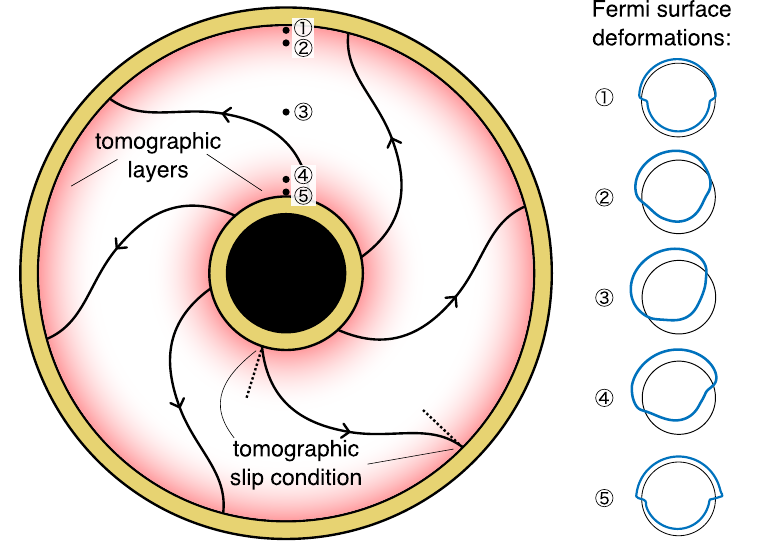}
    \caption{
    Left: Schematic of transport in a Corbino disk with electrodes shown in gold. Tomographic boundary layers are shaded red. Streamlines (black lines) emanate from the electrodes at a nonzero angle from the radial direction (dotted) due to the velocity slip conditions [Eq.~\eqref{eq:uBt2}]. Right: Exaggerated Fermi-surface deformations across the flow domain, with thin black circles indicating the equilibrium Fermi surface.
    }
    \label{fig:1}
\end{figure}

In high-mobility devices, electron transport becomes limited by electron-electron scattering when the electronic mean free path~$\ell_e$ becomes shorter than the device size~$L$~\cite{narozhny17,fritz24,hui25}. These flows are often described using a hydrodynamic model, where electronic scattering enters via the viscosity \mbox{$\eta=v_F \ell_e/4$} ($v_F$ is the Fermi velocity). Although hydrodynamic effects in interacting Fermi liquids have been observed~\citep{krishnakumar17,gooth18,berdyugin19,gupta21,madhogaria26,sulpizio19,bandurin18,bandurin16,kumar22}, a quantitative confirmation of the Fermi-liquid prediction for the viscosity has been surprisingly elusive: Experimental measurements show a considerable variation in the observed density and temperature scaling \mbox{$\eta \sim n^{0.5-3}$} and \mbox{$\eta \sim 1/T^{0.5-2}$}~\cite{krishnakumar17,gooth18,berdyugin19,gupta21,madhogaria26}, at variance with Fermi-liquid theory \mbox{$\eta_{\rm FL} \sim v_F^2 E_F/T^2 \sim n^2/T^2$} (\mbox{$n^{1/2}/T^2$} for monolayer graphene). A key challenge is to separate hydrodynamic transport from disorder scattering and scattering off device boundaries (which range from specular~\citep{bandurin16,torre15,lee16,gupta21,keser21} to diffuse reflection~\citep{sulpizio19,bandurin18,berdyugin19}). Here, the Corbino disk [cf.~Fig.~\ref{fig:1}] has emerged as a natural geometry that circumvents uncertainty in the boundary scattering~\cite{tomadin14,shavit19,holder19,zeng19,kumar22,vijayakrishnan23}: All device boundaries are electrode contacts where current is injected or drained. Recently, \textcite{zeng24} used this geometry and the suppression of the electron viscosity by a magnetic field to measure hydrodynamic and Ohmic transport in the same device, thus isolating the hydrodynamic contributions to transport. Yet, even here, a comparison with a hydrodynamic model reveals an anomalous temperature and density scaling of the extracted electron viscosity~\cite{zeng24}.

Recent efforts to explain these deviations from Fermi-liquid scaling have focused on the peculiarities of electron-electron scattering at low temperatures~\cite{kryhin25,rostami25}, which is known to differ markedly from the hydrodynamic picture that holds at intermediate temperatures: Pauli blocking gives rise to long-lived modes that are not relaxed efficiently by such scattering~\cite{gurzhi95,ledwith19,nilsson25}. This so-called tomographic effect is particularly pronounced for spin-symmetric modes in~2D but also exists in~3D~\cite{musser26} and for spin-imbalanced modes~\cite{raichev25} as well as in charge-neutral Fermi liquids~\cite{maki25}. Tomographic flows have been predicted to exhibit a fractional wavelength-dependent response in steady~\cite{ledwith19} and unsteady~\citep{thuillier26} flows, a distinct magnetoresponse~\cite{moiseenko24,rostami25}, and a modified low-temperature behavior~\cite{kryhin25,starkov25,estrada25}. Accurate modeling of flows in confined geometries is required to isolate tomographic flow phenomena, which can sensitively depend on boundary scattering and the geometry of the edges~\cite{benshachar25a,benshachar25b,starkov25,estrada25,farrell26}. However, a theory for tomographic transport driven by electrode contacts is currently largely missing. 

In this Letter, we derive and solve a transport theory for tomographic flows in a Corbino disk, obtained using a matched asymptotic expansion of the underlying kinetic equation. This reveals an extended ``tomographic boundary layer'' near each electrode, in which the system is far from local equilibrium and thus non-hydrodynamic. We find that the tomographic effect induces superballistic current injection at electrodes, which saturates an upper bound for the electrode conductance~\cite{raichev22}, as well as an anomalously large slip in the azimuthal current, which increases the Corbino disk's resistance. The density and temperature dependence that these tomographic corrections induce in the Corbino magnetoresistance may account for the observed disparate scaling in recent measurements. Moreover, the tomographic regime crosses over to the hydrodynamic regime at weak magnetic fields when the cyclotron radius falls below the (large) odd-mode mean free path. This phenomenon should permit extraction of the hydrodynamic viscosity and identification of the odd-mode mean free path from magnetoresistance measurements such as those performed in~\cite{zeng24}. 

We begin with the stationary Fermi-liquid equation for the quasiparticle distribution function $f$, written here in polar coordinates,
\begin{align}
  \biggl[v \sin\theta \frac{\partial}{\partial r} - \frac{v \cos\theta}{r} \frac{\partial}{\partial \theta} - \omega_c \frac{\partial}{\partial \theta}
\biggr] f(r, v, \theta) &= \mathcal{I}[f] ,
     \label{eq:BTE}
\end{align}
where $r$ is the radial position, and $v$ and $\theta$ denote the magnitude of the quasiparticle velocity and its orientation relative to the coordinate tangent [i.e., \mbox{$\theta=0$} points in the clockwise direction]. We include a constant out-of-plane magnetic field of strength~$B$ with cyclotron frequency \mbox{$\omega_c=eB/m^*$} ($m^*$ is the effective mass and \mbox{$e>0$} is the fundamental charge). An external electric potential is included via boundary conditions at the electrodes. $\mathcal{I}[\delta f]$ is the collision integral that describes the relaxation of the distribution to local equilibrium. This proceeds either by momentum-relaxing (MR) collisions to a stationary Fermi-Dirac equilibrium function, or by momentum-conserving (MC) collisions to a shifted Fermi-Dirac distribution. Both processes conserve the local chemical potential $\mu(r)$ (i.e., the electron number) and the latter also conserves the average velocity $u(r)$ (i.e., the momentum). For small flow velocities, we linearize the collision integral by expanding \mbox{$f \approx f_0 + \delta f$}, with $f_0$ the global Fermi-Dirac equilibrium distribution and a deviation
\begin{align}
    \delta f(r, v, \theta) &= E_F \biggl(- \frac{\partial f_0}{\partial \varepsilon}\biggr) \text{Ma} \, h(r, \theta) ,
\end{align}
where the Mach number is \mbox{$\text{Ma}=U/v_F$}, and \mbox{$U=I/(n e \ell_e)$} is a hydrodynamic velocity scale induced by a total injected charge current~$I$, with $n$ the electron density. Likewise, the linearized collision integral is defined as \mbox{$\mathcal{I}[\delta f] = (- \frac{\partial f_0}{\partial \varepsilon}) \mathcal{L}[h]$}. At low temperatures, $h$ is peaked close to the Fermi surface such that \mbox{$v \simeq v_F$}. The remaining dependence on the polar angle~$\theta$ is expanded as
\begin{align}
    h(r, \theta) &= \sum_{m \geq 0} \Bigl( a_m(r) \cos(m\theta) + b_m(r) \sin(m\theta) \Bigr) . \label{eq:hexp}
\end{align}
Rotational symmetry of the circular Fermi surface dictates that the collision integral is diagonal in the angular index $m$, with eigenvalues \mbox{$\gamma_m = \gamma_m^{\rm MR} + \gamma_m^{\rm MC}$}. We impose a fixed-relaxation-time model for MR scattering \mbox{$\gamma_{m\neq 0}^{\rm MR} =  \gamma^{\rm MR}$}, and parametrize MC relaxation rates based on exact-diagonalization studies~\cite{hofmann23,nilsson25} with a constant relaxation rate $\gamma_e$ for even-parity modes (even $m\geq 2$) and an odd-parity (odd $m\geq 3$) relaxation rate \mbox{$(\gamma_m^{\rm MC})^{-1} = (\gamma' m^p)^{-1} + \gamma_e^{-1}$} with \mbox{$\gamma' \ll \gamma_e$}. This interpolates between long-lived modes for small~$m$ and the Fermi-liquid form~$\gamma_e$ at large~$m$. The disparity between the relaxation rates of even and odd modes is the tomographic effect. Exact diagonalization studies indicate \mbox{$p=4$}~\cite{ledwith19b,nilsson25}, whereas the commonly employed constant odd-mode relaxation is obtained by setting \mbox{$p=0$}.

The Fermi liquid exhibits hydrodynamic flow if both momentum-conserving mean free paths \mbox{$\ell_e = v_F/\gamma_e$} and \mbox{$\ell_o=v_F/\gamma^{\rm MC}_3$} are shorter than the device dimension~$L$ (i.e., \mbox{$\ell_e,\ell_o \ll L$}), and if momentum-relaxing collisions are infrequent (i.e., \mbox{$L \ll \ell_{\rm MR}= v_F/\gamma_{\rm MR}$}). Here and in the following, we will take $L$ to be the distance between the inner and outer Corbino electrodes, i.e., \mbox{$L = r_2-r_1$}. As described earlier, in the presence of long-lived odd modes, where \mbox{$\ell_o>L$}, the hydrodynamic assumption does not hold and the system is said to be in the tomographic regime. A description of this regime is obtained here via a systematic expansion of the Fermi-liquid equation around the hydrodynamic limit~\cite{benshachar25a}. To this end, we define the even- and odd-mode Knudsen numbers as well as the Gurzhi number,
\begin{align}
    k_e=\frac{\ell_e}{L}, \quad k_o= \frac{\ell_o}{L}, \quad G = \frac{\sqrt{\ell_e \ell_{\rm MR}}}{L} ,
\end{align}
respectively. The above-mentioned length-scale separation requires \mbox{$k_e \ll 1$} while \mbox{$k_o, G \sim 1$}. We expand the distribution function in powers of~$k_e$ up to linear order,
\begin{align}
    h &= h^{(0)} + k_e h^{(1)} ,
    \label{eq:keExpansion}
\end{align}
as well as its projections
\begin{subequations}
\begin{align}
    u_r(r) &= \int_{-\pi}^\pi \frac{d\theta}{2\pi} h(r, \theta) \sin \theta = k_e u_r^{(1)}(r) \\
    u_\theta(r) &= \int_{-\pi}^\pi \frac{d\theta}{2\pi} h(r, \theta) \cos \theta = u_\theta^{(0)}(r) + k_e u_\theta^{(1)}(r) \\
    \delta \mu(r) &= \int_{-\pi}^\pi \frac{d\theta}{2\pi} h(r, \theta) = \delta \mu^{(0)}(r) + k_e \delta \mu^{(1)}(r) ,
\end{align}
\end{subequations}
where we anticipate that the radial velocity will start at linear order in $k_e$. Here, the velocity components $u_i$ are scaled by \mbox{$I/(e n\ell_e)$} and the potential \mbox{$\delta \mu$} by \mbox{$ e I/(2G_{\rm hydro})$}, where we define the hydrodynamic conductance scale \mbox{$G_{\rm hydro}=ne^2\ell_e/(m^* v_F)$}. Substituting into Eq.~\eqref{eq:BTE} and collecting powers of $k_e$ gives bulk equations for the macroscopic variables at each order of the expansion. These consist of continuity and Stokes-Ohm-like equations, with the latter modified by tomographic effects compared to its conventional hydrodynamic form. This expansion is independent of the flow domain boundaries and thus identical to the bulk expansion in~\cite{benshachar25a}. For brevity, we omit details of this derivation and summarize the results up to \mbox{${\it O}(k_e)$}. First, incompressibility dictates
\begin{align}
    \frac{\partial}{\partial r} \bigl( r u_{B|r}^{(1)} \bigr) &= 0 , \label{eq:divU}
\end{align}
where the subscript ``B" indicates bulk quantities, and~$r$ has been scaled by $L$. This gives a radial bulk flow \mbox{$u_{B|r}^{(1)} = 1/(2\pi r)$}. Second, the azimuthal projection of the generalized Stokes-Ohm equations describes a balance of the Lorentz force induced by the radial flow with viscous and Ohmic forces,
\begin{align}
    \frac{\Delta_r}{4} 
    u_{B|\theta}^{(0)} - \frac{1}{G^2} u_{B|\theta}^{(0)} &= - \frac{1}{r_c} u_{B|r}^{(1)} , \label{eq:uBtheta0}
\end{align}
where $\Delta_r$ is the radial component of the 2D Laplace operator, and \mbox{$r_c = v_F/\omega_c L$} is the dimensionless cyclotron radius. A finite-wavelength correction to this force balance arises at~\mbox{${\it O}(k_e)$},
\begin{align}
    &\frac{\Delta_r}{4} u_{B|\theta}^{(1)} - \frac{1}{G^2} u_{B|\theta}^{(1)} = \frac{1}{G^2} \frac{k_o}{(1+(3k_o/r_c)^2 )}\biggl[\frac{u_{B|\theta}^{(0)}}{G^2} - \frac{u_{B|r}^{(1)}}{r_c}\biggr] . \label{eq:uBtheta2}
\end{align}
Third, the radial projections of the generalized Stokes-Ohm equations dictate the local field,
\begin{subequations}
\begin{align}
    &\frac{\partial \delta \mu^{(0)}_B}{\partial r}  = - \frac{2}{r_c} u_{B|\theta}^{(0)} , \label{eq:deltamuB0} \\
    &\frac{\partial \delta \mu^{(1)}_B}{\partial r} = - \frac{2}{r_c} u_{B|\theta}^{(1)} - \frac{2}{G^2} u_{B|r}^{(0)} . \label{eq:deltamuB1}
\end{align}
\end{subequations}
Microscopically, this bulk description [Eqs.~\eqref{eq:divU},~\eqref{eq:uBtheta0} and~\eqref{eq:deltamuB0}] corresponds to a local equilibrium at leading order in the expansion (see also Eq.~\eqref{eq:hB0} below and panel~3 of Fig.~\ref{fig:1}). At $O(k_e)$ [Eqs.~\eqref{eq:uBtheta2} and~\eqref{eq:deltamuB1}], additional \mbox{$m=2$} and \mbox{$m=3$} modes arise (see Eq.~\eqref{eq:hB1}), which are associated with the electron viscosity and finite-wavelength tomographic correction, respectively. 

The bulk expansion alone is not compatible with microscopic boundary conditions at electrodes: Near each electrode the flow is inherently far from equilibrium. For perfect electrodes, 
where quasiparticles are transmitted from the leads, we impose for the injected quasiparticles
\begin{align}
    \bigl[ h(\theta \gtrless 0) \bigr]_S &= \pm \delta \mu_l , \label{eq:bc}
\end{align}
where \mbox{$[...]_S$} indicates evaluation at the surface of the flow domain (i.e., at the electrodes), and the top (bottom) sign applies at the inner (outer) electrode. Tomographic boundary layer corrections $h_{T,1/2}$, which describe the non-equilibrium dynamics in this region, must be added to the bulk solution to satisfy
Eq.~\eqref{eq:bc}:
\begin{align}
    h(r, \theta) &= h_B(r, \theta) + h_{T,1}(\chi_1, \theta) + h_{T,2}(\chi_2, \theta) ,
    \label{eq:hBT_sum}
\end{align}
where \mbox{$\chi_1=(r-r_1)/\sqrt{k_e k_o}$} and \mbox{$\chi_2=(r_2-r)/\sqrt{k_e k_o}$} are rescaled radial coordinates at the inner and outer electrodes. In conventional kinetic theory, the analogous boundary layers decay over distances comparable to the shortest mean free path $\ell_e$~\citep{benshachar25c,benshachar25d}. By contrast, here these corrections decay over an anomalously large distance scale \mbox{$\sqrt{\ell_e \ell_o}$} due to a combination of head-on scattering with short mean free path $\ell_e$ and small-angle scattering with long mean free path $\ell_o$. The solution for $h_T$ gives non-hydrodynamic corrections in the tomographic layer and dictates the boundary conditions for the bulk equations, which we discuss now. 

At \mbox{${\it O}(k_e^0)$}, we obtain the conventional hydrodynamic no-slip condition on the bulk azimuthal velocity, \mbox{$[u_{B|\theta}^{(0)}]_S = 0$}. A velocity slip condition arises at \mbox{${\it O}(k_e)$},
\begin{align}
    \label{eq:uBt2}
    &\Bigl[u_{B|\theta}^{(1)} \Bigr]_S = \pm \frac{32}{15\pi} \biggl[ \frac{\partial u_{B|\theta}^{(0)}}{\partial r} \biggr]_S  \nonumber \\ 
    & + \frac{k_o}{2(1+(3k_o/r_c)^2)} \biggl({\frac{3}{2}} \biggl[\frac{1}{r}\frac{\partial u_{B|\theta}^{(0)}}{\partial r} \biggr]_S - \frac{1}{2}\biggl[\frac{\partial^2 u_{B|\theta}^{(0)}}{\partial r^2} \biggr]_S \biggr) ,
\end{align}
which is identical to the condition derived in~\cite{benshachar25a} at a diffuse boundary. The first term on the right-hand side corresponds to a conventional slip-length condition with dimensional slip length \mbox{$\ell_{\rm slip}=(32/15\pi)\ell_e$}, while subsequent terms are driven by the tomographic effect (note the prefactor of~$k_o$). These dictate the dominant tomographic flow phenomena in a Corbino disk (indicated in Fig.~\ref{fig:1} as ``tomographic slip condition''), and are  suppressed for \mbox{$k_o/r_c \simeq 1$}, i.e., when the dimensional cyclotron radius is comparable to the odd-mode mean free path. Note that the tomographic slip condition depends on the curvature of the electrodes [the \mbox{$1/r$} term in Eq.~\eqref{eq:uBt2}], which permits tuning its magnitude with the device geometry: the smaller the (inner) Corbino radius, the larger the effect. Tomographic layer corrections to the azimuthal velocity and the gradient of the voltage at $O(k_e)$ are given in the End Matter.

The injection or drainage of current across the electrode is driven by a voltage drop from the electrode to the flow domain. We find that their relation is captured by a boundary condition on the radial velocity
\begin{align}
    \Delta \mu_{S} = \pm \delta \mu_l -\bigl[\delta \mu_B^{(1)}\bigr]_S = \pm \frac{\pi}{2} \bigl[&u_{B|r}^{(1)}\bigr]_S.
    \label{eq:uBn2}
\end{align}
Equation~\eqref{eq:uBn2} dictates a dimensionless conductance of each electrode \mbox{$G_{\rm e}=2[u_{B|r}^{(1)}]_S/(\Delta \mu_{S}k_e) = 4/\pi k_e$} which saturates an upper bound derived by~\textcite{raichev22} of twice the Sharvin conductance \mbox{$G_{\rm sh} = k_e^{-1}\int_0^\pi \frac{d\theta}{\pi} \sin \theta = 2/\pi k_e$} (this is in units of \mbox{$G_{\rm hydro} \sim k_e$}, such that the dimensional Sharvin conductance is independent of $\ell_e$). This electrode conductance can be understood as follows: Electrons entering the flow domain from the electrode provide a conductance contribution equal to the Sharvin conductance. After traveling a short distance $\ell_e$, they undergo a head-on scattering event, which prevents a second electron (their partner in the scattering) from entering the electrode. As described by~\textcite{ginzburg23,hong24}, this scattering can be modeled as holes back-propagating along the electron trajectories in reverse into the electrode, which gives a further Sharvin conductance contribution to the electrode conductance. The corresponding microscopic distribution is illustrated in the bottom panel of Fig.~\ref{fig:1}, where the distribution of velocities entering the electrode is the negative of the outgoing electrons relative to the local potential (i.e., describing retroreflected holes). An analogous process occurs at the outer electrode (see top panel of Fig.~\ref{fig:1}), where holes are injected into the flow domain, and are retroreflected as electrons back into the electrode. Finite mean-free path and curvature corrections to the electrode conductance~\citep{nagaev2018,uzair2018,raichev22} do not arise at this order of expansion.

\begin{figure}
    \centering
    \includegraphics[width=\linewidth]{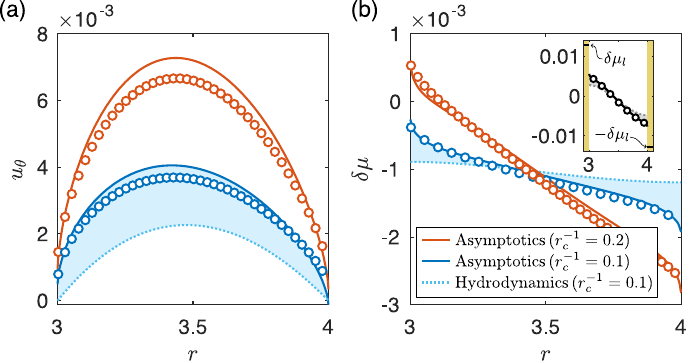}
    \caption{Profiles of the (a) azimuthal velocity and (b) electrochemical potential in the Corbino disk with \mbox{$G\to \infty$}, \mbox{$k_e=0.1$}, \mbox{$k_o=3$}, \mbox{$p=0$}, \mbox{$r_1=3$} and \mbox{$1/r_c=0.1$} (blue) and \mbox{$1/r_c=0.2$} (red). Open circles are numerical solutions of Eq.~\eqref{eq:BTE} with \mbox{$p=0$} and solid lines are the asymptotic theory. The hydrodynamic solution is shown for \mbox{$1/r_c=0.1$} (dotted line), and the difference between this and the asymptotic theory is shaded. The inset in (b) shows the full voltage profile for $1/r_c=0.5$ including the potential in the electrodes (shaded gold) at $\pm \delta \mu_l$.
    }
    \label{fig:2}
\end{figure}

The asymptotic solution up to $O(k_e)$ is obtained by solving Eq.~\eqref{eq:uBtheta0} with no-slip boundary conditions, then solving Eq.~\eqref{eq:uBtheta2} with the tomographic boundary condition~\eqref{eq:uBt2}, integrating Eqs.~\eqref{eq:deltamuB0} and~\eqref{eq:deltamuB1} for the bulk voltage drop, and finally adding the tomographic layer corrections in Eqs.~\eqref{eq:uT2} and~\eqref{eq:dmuT} to the bulk solutions (see End Matter). The total voltage drop across the electrodes is obtained by adding the drop at the electrodes, \mbox{$2\mu_l=\int_{r_1}^{r_2} dr \, \partial_r \delta \mu_B(r) + \Delta \mu_{1} + \Delta \mu_{2}$}, which then gives the resistance \mbox{$R=\mu_l$} (in units of $G_{\rm hydro}^{-1}$).

Figure~\ref{fig:2} shows the azimuthal velocity~$u_\theta(r)$ and the voltage~$\delta \mu(r)$ for \mbox{$k_e=0.1$}, \mbox{$k_o=3$}, \mbox{$r_1=3, G\to\infty$} for two magnetic field strengths \mbox{$1/r_c = 0.1$} and \mbox{$1/r_c = 0.2$}. To benchmark our results, we overlay direct numerical solutions of the kinetic equation~\eqref{eq:BTE} for \mbox{$p=0$} (open circles). We also show the hydrodynamic no-slip solution for \mbox{$1/r_c=0.1$} (dotted line), such that the shaded region in Fig.~\ref{fig:2} indicates the tomographic correction. Even for relatively small $k_e$, the hydrodynamic theory underpredicts the true solution by more than a factor of two, while the error between the asymptotic theory and the numerical solution is at most~9\%. The strong enhancement of the azimuthal flow is due to the tomographic slip condition~\eqref{eq:uBt2}, which, via the Lorentz force, produces a commensurate enhancement in the voltage drop across the electrodes. For illustration, the inset in Fig.~\ref{fig:2}(b) shows the full voltage distribution for \mbox{$1/r_c=0.5$} including the drop between the electrodes and the flow domain. 

\begin{figure}
    \centering
    \includegraphics[width=\linewidth]{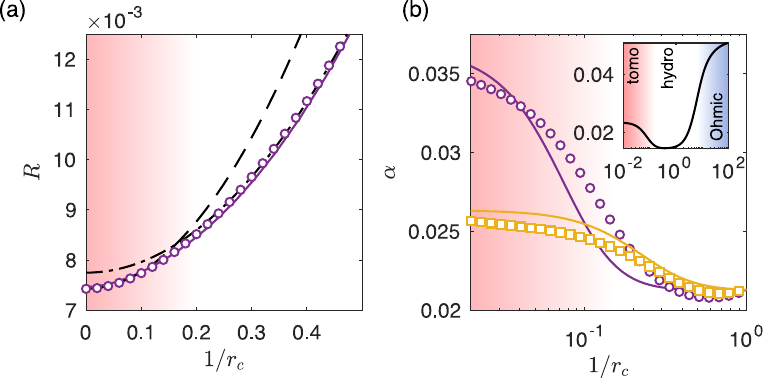}
    \caption{
    (a) Resistance as a function of the magnetic field strength for \mbox{$G\to \infty$}, \mbox{$k_e=0.1$}, \mbox{$k_o=3$}, \mbox{$r_1=3$}. Open circles show numerical solutions. The solid purple line is the asymptotic solution (shifted by a constant). The dashed line shows the tomographic magnetoresistance curvature that applies at weak magnetic fields, and the dash-dotted line shows the hydrodynamic result valid at higher fields. (b) Quadratic magnetoresistance coefficient \mbox{$\alpha$} for \mbox{$k_o=3$} (purple) and \mbox{$k_o=1$} (yellow). The main panel highlights the  crossover from tomographic to hydrodynamic scaling  at weak fields \mbox{$1/r_c\simeq 1/k_o$}. Inset: Full magnetic-field dependence of $\alpha$ for  \mbox{$G=1,k_o=3$}, including the Ohmic regime at large magnetic fields. The tomographic, hydrodynamic, and Ohmic regimes are shaded red, white, and blue, respectively.
    }
    \label{fig:3}
\end{figure}

Figure~\ref{fig:3}(a) shows the magnetoresistance as a function of the (dimensionless) magnetic-field strength \mbox{$1/r_c$}, which is described by a parabolic field dependence (illustrated by the dashed and dash-dotted lines) with locally varying curvature
\begin{align}
    \alpha\bigl(1/r_c\bigr) = \frac{\partial R}{\partial (r_c^{-2})} ,
    \label{eq:alpha}
\end{align}
whose value is characteristic of the transport regime; $\alpha$ is larger in the Ohmic regime than the hydrodynamic regime~\citep{zeng24}. In the tomographic regime, the enhanced velocity slip dictates that $\alpha$ is larger than its hydrodynamic value. This slip, however, is rapidly suppressed at weak magnetic fields when \mbox{$r_c \simeq k_o$}, with a corresponding suppression of $\alpha$, and the hydrodynamic regime is recovered. The predicted tomographic suppression of $\alpha$ is in excellent agreement with direct numerical solutions of the kinetic equation; see Fig.~\ref{fig:3}(b). Thus, with finite momentum relaxation, three regimes emerge: a tomographic regime at zero and weak magnetic fields (i.e., for \mbox{$k_o \lesssim r_c<\infty$}), a hydrodynamic regime at intermediate fields (i.e., \mbox{$k_e \lesssim r_c \lesssim k_o$}), and an Ohmic regime at large magnetic fields (i.e., \mbox{$r_c \lesssim k_e$}) when the viscosity is suppressed. This implies a non-monotonic dependence of $\alpha$ on the magnetic field, which first decreases at weak fields and then increases at large fields. The main panel of Fig.~\ref{fig:3}(b) highlights the tomographic-to-hydrodynamic crossover, and the inset of Fig.~\ref{fig:3}(b) shows the full magnetic-field dependence including the Ohmic regime for \mbox{$G=1$}, where a prefactor \mbox{$(1+(2k_e/r_c)^2)^{-1}$} is added to the first term of Eqs.~\eqref{eq:uBtheta0} and~\eqref{eq:uBtheta2} for the magnetic-field suppression of the viscosity.

Exact diagonalization studies of the collision integral~\citep{ledwith19b,nilsson25} predict an asymptotic Fermi-liquid scaling \mbox{$k_e\sim n^{3/2}/T^2$} for the even-mode Knudsen number, while the odd-mode Knudsen number has a distinct scaling \mbox{$k_o\sim n^{7/2}/T^4$}. This sets the scaling behavior of the magnetoresistance curvature. Fitting $\alpha$ at weak fields with a viscous response following~\citep{zeng24} gives an associated electronic timescale $\tau_\eta$: The leading-order (hydrodynamic) prediction gives the usual Fermi-liquid scaling \mbox{$\tau_\eta^{-1} \sim T^2/n$}, while the leading-order tomographic correction has opposite temperature and density scaling \mbox{$(\tau_\eta')^{-1} \sim n^4/T^4$}, with an additional (non-tomographic) finite-size contribution that scales as \mbox{$(\tau_\eta'')^{-1}\sim \sqrt{n}$}. Note that the scaling corrections considered here arise from a detailed treatment of electrode boundary effects and are thus distinct from those discussed in~\citep{kryhin25,rostami25}, where bulk flows were considered. The full temperature and density dependence arises from a sum of these terms---whose relative magnitudes can be tuned with the curvature of the electrodes---suggesting that the tomographic corrections at weak magnetic fields could be responsible for the anomalous temperature scalings observed in experiments, where the weak-field dependence is attributed to viscosity alone~\citep{zeng24}. 

Our results indicate that a more robust measurement of the electron viscosity would be obtained in these comparisons by replacing the value of $\alpha$ at weak magnetic fields with its value at moderate magnetic fields, where the tomographic regime is suppressed. Beyond a direct measurement of the electron viscosity, the non-monotonic dependence of $\alpha$ on the magnetic field strength should serve as a signature of the tomographic effect, and, if observed, would give estimates for both the odd- and even-mode mean free paths.

\begin{acknowledgments}
We thank Jeff Maki for discussions. This work is supported by Vetenskapsr\aa det (Grant No.~2024-04485), the Olle Engkvist Foundation (Grant No.~233-0339), the Knut and Alice Wallenberg Foundation (Grant No.~KAW 2024.0129), and Nordita.
\end{acknowledgments}

\bibliography{bib_corbino}

\appendix
\section{End Matter}

Here, we collect expressions for the tomographic boundary layer functions, which need to be added to the bulk solutions to be consistent with microscopic boundary conditions. We also list explicit expressions for the bulk solutions in the Corbino disk. 

\subsection{Matched asymptotic expansion}

The bulk distribution function up to $O(k_e)$ is
\begin{subequations}
\label{eq:bulkdeformation}
\begin{align}
    h_B^{(0)} &= \delta \mu^{(0)}_B - 2 \cos \theta \, u_{B|\theta}^{(0)} \label{eq:hB0}\\
    h_B^{(1)} &= \delta \mu^{(1)}_B + 2 \sin \theta \, u_{B|r}^{(1)} - 2 \cos \theta \, u_{B|\theta}^{(1)} \nonumber \\
    &\quad + \sin 2 \theta \biggl(\frac{\partial u_{B|\theta}^{(0)}}{\partial r} - \frac{u_{B|\theta}^{(0)}}{r}\biggr) \nonumber \\
    &\quad + \frac{k_o}{2 (1+(3k_o/r_c)^2)} \biggl[\cos 3\theta + \frac{3k_o}{r_c} \sin 3 \theta\biggr] \nonumber \\
    &\qquad \times r^2 \frac{\partial}{\partial r} \biggl[\frac{1}{r^2} \biggl(\frac{\partial u_{B|\theta}^{(0)}}{\partial r} - \frac{u_{B|\theta}^{(0)}}{r}\biggr)\biggr] . \label{eq:hB1}
\end{align}
\end{subequations}
As discussed in the main text, the leading-order $h_B^{(0)}$ [Eq.~\eqref{eq:hB0}] takes a hydrodynamic form that consists of density and current deformations. The next-to-leading-order term $h_B^{(1)}$ also includes an even-parity \mbox{$m=2$} admixture that indicates viscous shear flow [second line of Eq.~\eqref{eq:hB1}] as well as long-lived odd-parity deformations with \mbox{$m=3$} [third line].

\begin{figure}[t!]
    \centering
    \includegraphics[width=\linewidth]{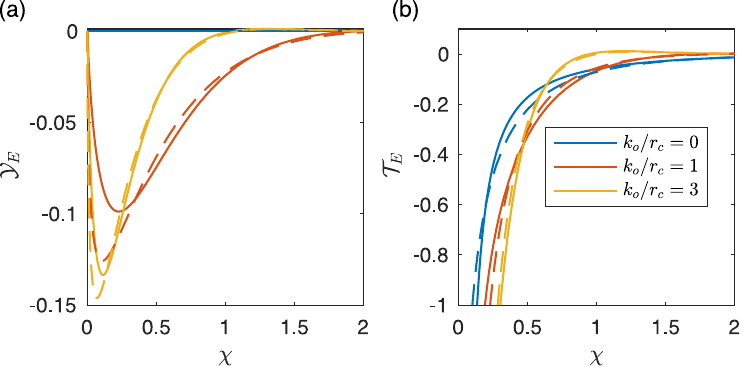}
    \caption{
    Tomographic layer functions (a) \mbox{${\cal Y}_E$} and (b) \mbox{${\cal T}_E$} for \mbox{$p=4$} and \mbox{$k_o/r_c\in\{0,1,3\}$}, with  \mbox{$\gamma'/\gamma_e=0$} (solid lines) and \mbox{$\gamma'/\gamma_e=1$} (dashed lines).}
    \label{fig:4}
\end{figure}

Substituting Eqs.~\eqref{eq:hBT_sum} and~\eqref{eq:keExpansion} into Eq.~\eqref{eq:BTE} and collecting powers of $\sqrt{k_e}$ gives the governing equation for the leading-order tomographic correction,
\begin{align}
    &\sqrt{k_o}\sin\theta  \frac{\partial \delta \mu_T^{(3/2)}}{\partial \chi} = \sin^2\theta \frac{\partial^2 h_T^{(1)}}{\partial \chi^2} - L\bigl[h_T^{(1)}\bigr]  - \frac{k_o}{r_c} \frac{\partial h_T^{(1)}}{\partial \theta}, \label{eq:tomographiceq}
\end{align}
where we denote the order by $(3/2)$ since it formally arises as an ${\it O}(k_e^{3/2})$ term in the expansion~\cite{benshachar25a}. The same substitution also constrains \mbox{$h_T^{(1)}$} to have odd parity and furthermore  \mbox{$u_{T|r}^{(1)} = 0$} (i.e., the radial current is determined by the incompressibility condition without additional tomographic inflow near the electrode). Equation~\eqref{eq:tomographiceq} is solved subject to the boundary conditions
\begin{subequations}
\begin{align}
    h_T^{(1)}\Bigr|_{\chi=0} &= \pm \mu_l - h_B^{(1)}\Bigr|_{\chi=0}, \label{eq:BC1_tomo} \\
    \lim_{\chi \to \infty} \chi^a h_T^{(1)} &= 0 , \label{eq:BC2_tomo}
\end{align}
\end{subequations}
for all powers $a$. Substituting the bulk distribution~\eqref{eq:bulkdeformation} into Eq.~\eqref{eq:BC1_tomo} and enforcing \mbox{$u_{T|r}^{(1)} = 0$} gives the penetration condition in Eq.~\eqref{eq:uBn2}, while enforcing Eq.~\eqref{eq:BC2_tomo} provides the slip condition~\eqref{eq:uBt2}. The solution for the tomographic corrections then takes the form 
\begin{widetext}
\begin{subequations}
\begin{align}
        u_{T|\theta}^{(1)} &= 
        - {\cal Y}_E\Bigl(\chi; \frac{k_o}{r_c}\Bigr) \Bigl[ u_{B|r}^{(1)}\Bigr]_S 
        \mp \mathcal{Y}_0\Bigl(\chi;\frac{k_o}{r_c}\Bigr) \biggl[\frac{\partial u_{B|\theta}^{(0)}}{\partial r} \biggr]_S 
        -{\frac{3 k_o}{2}} \mathcal{Y}_1\Bigl(\chi;\frac{k_o}{r_c}\Bigr) \biggl[\frac{1}{r} \frac{\partial u_{B|\theta}^{(0)}}{\partial r} \biggr]_S
        + \frac{k_o}{2} \mathcal{Y}_1\Bigl(\chi;\frac{k_o}{r_c}\Bigr) \biggl[\frac{\partial^2 u_{B|\theta}^{(0)}}{\partial r^2} \biggr]_S , \label{eq:uT2} \\
        \frac{\partial \delta \mu_T^{(3/2)}}{\partial \chi} 
        &= 
        \pm \frac{1}{\sqrt{k_o}} {\cal T}_E\Bigl(\chi;\frac{k_o}{r_c}\Bigr) \Bigl[ u_{B|r}^{(1)}\Bigr]_S 
        \mp \frac{2\sqrt{k_o}}{r_c} u_{T|\theta}^{(1)} 
        + \frac{1}{\sqrt{k_o}}\mathcal{T}_0\Bigl(\chi;\frac{k_o}{r_c}\Bigr) \biggl[\frac{\partial u_{B|\theta}^{(0)}}{\partial r} \biggr]_S 
        \pm {\frac{3}{2}} \sqrt{k_o}\mathcal{T}_1\Bigl(\chi;\frac{k_o}{r_c}\Bigr) \biggl[\frac{1}{r} \frac{\partial u_{B|\theta}^{(0)}}{\partial r} \biggr]_S \nonumber \\
        & \quad \mp \frac{\sqrt{k_o}}{2} \mathcal{T}_1\Bigl(\chi;\frac{k_o}{r_c}\Bigr) \biggl[\frac{\partial^2 u_{B|\theta}^{(0)}}{\partial r^2}\biggr]_S , \label{eq:dmuT} \\
         h_T^{(1)} &= 
        f_E(\chi, \theta) \, \bigl[  u_{B|r}^{(1)} \bigl]_S 
        \pm  f_0\Bigl(\chi,\theta;\frac{k_o}{r_c}\Bigr) \biggl[\frac{\partial u_{B|\theta}^{(0)}}{\partial r} \biggr]_S
        + \biggl(\frac{3 k_o}{2}  \biggl[\frac{1}{r} \frac{\partial u_{B|\theta}^{(0)}}{\partial r} \biggr]_S 
        - \frac{k_o}{2} \biggl[\frac{\partial^2 u_{B|\theta}^{(0)}}{\partial r^2} \biggr]_S \biggr) f_1\Bigl(\chi,\theta;\frac{k_o}{r_c}\Bigr) . \label{eq:hT}
\end{align}
\end{subequations}
\end{widetext}
The last four terms on the right-hand side of Eq.~\eqref{eq:uT2} describe non-hydrodynamic corrections to the velocity profile driven by boundary stresses. They are identical to those derived in~\cite{benshachar25a} for diffuse reflection, and thus, for brevity, are not discussed further here. The first term in Eq.~\eqref{eq:uT2} is a correction driven by the injected velocity at the electrode, and thus does not arise in the diffuse analysis of~\cite{benshachar25a,benshachar25b}. As shown in Fig.~\ref{fig:4}(a), this correction vanishes in the absence of a magnetic field and can thus be interpreted as arising from the Lorentz force acting on injected electrons near the electrode. Similarly, the last five terms on the right-hand side of Eq.~\eqref{eq:dmuT} are identical to the diffuse reflection result, while the first term in Eq.~\eqref{eq:dmuT} is associated with injected current and is shown in Fig.~\ref{fig:4}(b). This correction is analogous to the correction given in Fig.~3 of~\cite{raichev22}. Substituting Eqs.~\eqref{eq:uT2}--\eqref{eq:hT} into Eq.~\eqref{eq:tomographiceq} and collecting the $[h_{B|r}^{(1)}]_S$ terms gives the required equation for $f_E$, subject to the boundary condition at $\chi=0$,
\begin{align}
    [f_E]_S = \frac{\pi}{2} \text{sgn}(\theta) + 2\sin\theta ,
\end{align}
which is solved using standard numerical techniques~\citep{benshachar25a}.

\subsection{Flow in Corbino disks}

Solving Eqs.~\eqref{eq:uBtheta0} and~\eqref{eq:uBtheta2} subject to the boundary conditions in Eq.~\eqref{eq:uBn2} and~\eqref{eq:uBt2} gives the azimuthal velocity
\begin{subequations}
\begin{align}
    u_\theta^{(0)} &= \frac{G^2}{2\pi r_c} \biggl[\frac{1}{r} + A_0 I_1\biggl( \frac{2r}{G}\biggr) + B_0 K_1 \biggl( \frac{2r}{G}\biggr) \biggr], \\
    u_\theta^{(1)} &= \frac{1}{2\pi} \biggl[ A_2 I_1\biggl( \frac{2r}{G}\biggr) + B_2 K_1 \biggl( \frac{2r}{G}\biggr) \biggr] + u_{{\rm FW}|\theta}^{(1)} + u_{T|\theta}^{(1)} , 
\end{align}
\end{subequations}
where $I_1$ and $K_1$ are modified Bessel functions of the first and second kind, with the constants
\begin{subequations}
\begin{align}
    A_0 &= \frac{1}{C} \biggl[\frac{1}{r_1} K_1\biggl(\frac{2r_2}{G}\biggr) - \frac{1}{r_2} K_1\biggl(\frac{2r_1}{G}\biggr)\biggr], \\
    B_0 &= - \frac{1}{C} \biggl[\frac{1}{r_1} I_1\biggl(\frac{2r_2}{G}\biggr) - \frac{1}{r_2} I_1\biggl(\frac{2r_1}{G}\biggr)\biggr] , \\
    A_2 &= -\frac{1}{C} \biggl[ K_1\biggl(\frac{2r_2}{G}\biggr) \bigl[u_{B|\theta}^{(1)} \bigr]_1 - K_1\biggl(\frac{2r_1}{G}\biggr) \bigl[u_{B|\theta}^{(1)} \bigr]_2\biggr] , \\
    B_2 &= \frac{1}{C} \biggl[ I_1\biggl(\frac{2r_2}{G}\biggr) \bigl[u_{B|\theta}^{(1)} \bigr]_1 - I_1\biggl(\frac{2r_1}{G}\biggr) \bigl[u_{B|\theta}^{(1)} \bigr]_2\biggr] ,
\end{align}
\end{subequations}
and
\begin{align}
    C = K_1\biggl(\frac{2r_1}{G}\biggr) I_1\biggl(\frac{2r_2}{G}\biggr) - I_1\biggl(\frac{2r_1}{G}\biggr) K_1\biggl(\frac{2r_2}{G}\biggr) .
\end{align}
The constants $A_2$ and $B_2$  contain the velocity slip conditions~\eqref{eq:uBt2} at the inner and outer electrodes 
\begin{align}
    \bigl[u_{B|\theta}^{(1)} \bigr]_{1,2} &= \frac{G^2}{2\pi r_c} \biggl[\mp \frac{32}{15 \pi} F(r) \nonumber \\&- \frac{k_o}{4 (1+(3k_o/r_c)^2) r} \Bigl(5 F(r) + H(r)\Bigr)\biggr]_{1,2}
\end{align}
with
\begin{align}
    F(r) &= \frac{1}{r^2} - \frac{A_0}{G} \biggl(I_0\biggl(\frac{2 r}{G}\biggr) + I_2\biggl(\frac{2 r}{G}\biggr)\biggr) \nonumber \\
    &\quad+ \frac{B_0}{G} \biggl(K_0\biggl(\frac{2 r}{G}\biggr) + K_2\biggl(\frac{2 r}{G}\biggr)\biggr) , \\
    H(r) &= \frac{2 A_0}{G^2} \biggl(2r I_1\biggl(\frac{2 r}{G}\biggr) + G I_0\biggl(\frac{2 r}{G}\biggr)\biggr) \nonumber \\ &\quad+ \frac{2 B_0}{G^2} \biggl(2 r  K_1\biggl(\frac{2 r}{G}\biggr) -G  K_0\biggl(\frac{2 r}{G}\biggr)\biggr) ,
\end{align}
where the top sign applies at the inner electrode (\mbox{$r=r_1$}) and the bottom sign at the outer electrode (\mbox{$r=r_2$}). The finite-wavelength correction \mbox{$u_{FW|\theta}^{(1)}$} at \mbox{${\it O}(k_e)$} is the inhomogeneous solution to Eq.~\eqref{eq:uBtheta2},
\begin{align}
    \frac{u_\theta^{(1)}}{G^2} &- \frac{\Delta_r}{4} 
    u_\theta^{(1)} = \frac{k_o}{(1+(3k_o/r_c)^2)} \nonumber \\& \times 
    \frac{1}{2\pi r_c G^2}\biggl[ A_0 I_1\biggl( \frac{2r}{G}\biggr) + B_0 K_1 \biggl( \frac{2r}{G}\biggr)\biggr] ,
\end{align}
with no-slip boundary conditions. This equation does not appear to have a closed-form solution but is easily solved numerically. We note that $u_{{\rm FW}|\theta}^{(1)}$ vanishes in the clean limit \mbox{$G\to\infty$}. Finally, the bulk voltage distribution $\delta \mu$ is obtained by integrating Eqs.~\eqref{eq:deltamuB0} and~\eqref{eq:deltamuB1}.

\end{document}